\documentclass[10pt,journal,compsoc]{IEEEtran}

\usepackage{amsmath}
\usepackage{cite}
\usepackage{color}
\usepackage[caption=false,font=normalsize,labelfont=sf,textfont=sf]{subfig}
\usepackage{stfloats}
\usepackage{hyperref}
\usepackage{multirow}
\usepackage{enumitem}
\usepackage{wrapfig}

\definecolor{lightgray}{rgb}{0.95,0.95,0.95}

\usepackage[pdftex]{graphicx}
\graphicspath{{figures/}}
\DeclareGraphicsExtensions{.pdf,.jpeg,.png}

\begin{document}

\title{Engineers Code: reusable open learning modules for engineering computations}
\author{Lorena A. Barba
\IEEEcompsocitemizethanks{\IEEEcompsocthanksitem Mechanical and Aerospace Engineering,
the George Washington University, Washington, DC 20052.\protect\\
Email: labarba@gwu.edu}
}

\IEEEtitleabstractindextext{%
\begin{abstract}
Undergraduate programs in science and engineering include at least one course in basic programming, but  seldom presented in a contextualized format, where computing is a tool for thinking and learning in the discipline. 
We have created a series of learning modules to embed computing in engineering education, and share this content under permissive public licenses. 
The modules are created as a set of lessons using Jupyter notebooks, and complemented by online courses in the Open edX platform, using new integrations we developed. 
Learning sequences in the online course pull content dynamically from public Jupyter notebooks and assessments are auto-graded on-the-fly, using our Jupyter Viewer and Jupyter Grader third-party extensions for Open edX (XBlocks). 
The learning content is modularized and designed for reuse in various formats. 
In one of these formats---short but intense workshops---our university library is leveraging the curriculum to offer extra-curricular training for all, at high demands.
\end{abstract}
}

\maketitle

\IEEEraisesectionheading{\section{Introduction}\label{sec:introduction}}

\IEEEPARstart{S}{cience} and engineering undergraduate programs routinely include in their curriculum a basic programming course, often provided as a service course by the local department of computer science. 
Regardless of the programming language used, one common approach for teaching non-CS (computer science) majors is to simplify the standard introductory CS course, to create a ``lightweight'' version focusing on programming basics. 
Less common---even though it is known to be effective---is to teach programming \emph{in context}. 
The classic example is the media-computing course introduced nearly two decades ago at the Georgia Institute of Technology (for liberal arts, architecture, and management/business majors) \cite{guzdial2003media,guzdial2005design}. 
Evaluation efforts on that multi-year curricular innovation support the idea that context-based teaching of programming increases student motivation and success \cite{forte2005motivation,guzdial2013exploring}. 
(In the Georgia Tech experience, it also reduced the success gender gap.) 
A more recent effort to introduce contextualized computing education in engineering found that it was effective in enabling students to apply computational practices to continue learning in their discipline \cite{magana2016case}. 
In view of their observations, the authors recommended integrating context-based computing early and often in the engineering curriculum. 

In this paper, I describe an initiative to develop a series of learning modules aimed at integrating computing in the undergraduate engineering curriculum. 
The modules adopt the context-based format for teaching programming, and are also designed to be reusable, and shared under standard public licenses (CC-BY for content and BSD-3 for code). 
We first developed three learning modules---each adding up to about one university credit of work---and taught a second-year engineering course based on this content  in Fall 2017 and Fall 2018. 
The faculty of the Mechanical and Aerospace Engineering department then approved creating a two-course series in computing, and we used the first two modules for a first-year course in Spring 2019, and started writing two additional learning modules to complete a revised second-year course. 
We are working with colleagues to develop additional learning modules to use within core engineering courses, aiming to reinforce the ability of students to use computational practices and problem-solving in their discipline. 
The design patterns we adopted (described in the next section) are the product of many years of teaching computational topics to engineering students, and adopting increasingly popular technologies (especially Jupyter notebooks). 
We use a hybrid-learning approach, combining an online course platform (Open edX) and auto-grading of student assignments, and active learning in the classroom meetings. 
This project fully embraces open-source tools and open educational practices, and its goal is to advance innovation in engineering education by integrating computing across the curriculum, disseminating our products broadly, and inviting collaboration.

\section{Key concepts and design principles}

Some key concepts and design principles in the \emph{Engineers Code} series of learning modules are: 
(1) the idea of ``computable content'': educational content made powerfully interactive via compute engines in the learning platform; 
(2) the idea of open pedagogy: reflecting in the teaching practice the ethos and practices of open source software; 
(3) modularization: creating stackable learning modules that break-up the standard ``course'' format; 
(4) harnessing the worked-example effect: empirically shown to be superior to problem-solving for novice learners; 
(5) using live-coding to structure active-learning class experiences; and
(6) guiding learners to document their own work. 

Our chosen learning platform is Jupyter: a browser-based interactive computing environment, concretized in a document format that seamlessly interleaves code with text-based and multi-media content: the Jupyter Notebook. 
Each learning module consists of four or five fully narrated Jupyter notebooks (the lessons), and student assignments also prepared and submitted as Jupyter notebooks.
I started using Jupyter for teaching in 2013 (when it still had not adopted this name). 
Based on a practical module used in the classroom in my Computational Fluid Dynamics (CFD) course (taught from 2010 to 2013 at Boston University), my first series of fully narrated notebooks is ``CFD Python: the 12 steps to Navier-Stokes equations'' \cite{BarbaForsyth2018}. 
Based on the experience creating and using the CFD Python learning module, and following a similar approach in later courses, we adopted this basic design pattern for developing lessons using computable content:

\begin{enumerate}
\item Break it down into small steps
\item Chunk small steps into bigger steps
\item Add narrative and connect
\item Link out to documentation
\item Interleave easy exercises
\item Spice with challenge questions/tasks
\item Publish openly online
\end{enumerate}

The \emph{Engineers Code} learning modules are published as Open Educational Resources (OER): anyone can access, reuse, revise, and redistribute the materials. 
The idea of creating educational materials that are made to be reused goes back twenty five years, and led to the development of content licenses like Creative Commons. 
Recurring topics in the conversations around OER are the high cost of textbooks, increasing access to content (for worldwide learners), questions of copyright and licenses, and values around altruism and the public good. 
But arguably OER have not been transformational: various surveys show that faculty for the most part still require that students purchase commercial textbooks, and have little awareness of OER alternatives. 
Even if the open education movement was inspired by open source software, it missed some key features: open development, networked collaboration, community, and a value-based framework. 
In open-source development, we cherish our \emph{productive freedom}: the freedom to work and collaborate by our own conventions, side-step the restrictions of copyright law by attaching a license to our products, and prioritize access, distribution, and collaboration. 
In OER development, the narrative is often about \emph{creation} of content, and \emph{adoption} by others. There is the Author, and there is the Adopter, or User. 
Although Creative-Commons licenses are meant for reuse and remix, in practice the emphasis is on sharing for reuse ``as is.'' 
(Consider for example the MIT Open Course Ware initiative: faculty create their course materials and deposit them for free access; users cannot become contributors.) 
Deliberately embracing the ethos and practice of open source software may not only lead to greater reuse, but could inspire students to more collaboration.

Modularization of the content is also aimed at increasing reuse. 
Course instructors all know how difficult it is to adopt another instructor's materials to teach a course, as we almost always would teach it differently. 
Science and engineering textbooks are often excessively long, and courses based on them leave out substantial portions. 
Each instructor may choose different chapters to leave out, while all students still must buy the bloated full book. 
Unbundling a university course into smaller content units empowers instructors to ``mix-and-match'' and adapt their course to local goals and student cohorts. 
It also reflects better how learners consume content today, and how users interact with all sorts of media on digital platforms. 
Modular content design also enables just-in-time use, whether in support of a longer course, or through informal learning opportunities (e.g., short tutorials, library workshops, peer learning groups). 

In the \emph{Engineers Code} series, each module is designed to comprise about one credit-hour of course work, and take about four or five weeks to complete in regular course scheduling. 
Each module, in turn, consists of a handful of lessons written as Jupyter notebooks. 
And each lesson develops a topic through a scaffolded design that takes advantage of the \emph{worked-example effect}: one of the best known and most widely studied of the cognitive load effects. 
The effect is positive when providing full guidance on how to solve a problem results in better student performance than problem-solving conditions with no guidance. 
Many studies have shown significant learning improvements with worked examples, compared with free-form problem-solving, for novices learning complex topics.
The opposite is called generation effect: 
when learners generating responses perform better than learners in a presentation setting that provides an answer. 
Chen et al.\ \cite{chen2015worked} concluded that the ``worked example effect occurs for complex, high-element interactivity materials that impose a heavy working memory load whereas the generation effect is applicable for low-element interactivity materials.''
Thus, strong guidance is best for novice learners dealing with complex materials; as students become more expert in the subject, guidance can be reduced in favor of discovery based learning. 
Needless to say, learning to program imposes a high cognitive load, suggesting that inexperienced trainees will benefit from worked-example effects. 
Each lesson in our collection aims to harness this effect by breaking down a computational problem into atomic steps, providing a detailed narrative and documentation of those steps, and peppering the narrative with low-stakes exercises for students. 

In the classroom, we adopt the method referred to in open source communities as \emph{live coding}. 
The instructor projects an interactive computing session (a draft Jupyter notebook), and demonstrates live the problem-solving sequence by typing and executing every command, while students follow along in their own interactive session. 
Naturally, mistakes happen, and the instructor has the opportunity to vocalize the corrections to these mistakes, modeling a process (debugging) that novices can be frustrated by when working alone. 
Students following along in class can mistype, or decide to try a small change, and encounter different errors during class. To help them, the instructor is aided by in-class learning assistants: undergraduate students who have recently completed the course. 
We use the familiar Software Carpentry ``Post-It Notes'' method: all students are handed one pink and one green sticky note. 
When they have a problem or a question, they stick the pink note on their computer monitor (and a learning assistant can come to them); when they are finished with a class exercise, they stick the green note on their monitor (and the instructor can get a feel for the room with one look around).
The live-coding approach, supported by in-class learning assistants, leads to an active learning format that works. 
Active learning is known to be superior to lectures both in terms of comprehension and recollection (memory) \cite{freeman2014active}.
Finally, students are encouraged to take their in-class notes in a new Jupyter notebook as a draft that they can continue adding to when they self-study. 
The goal is to model the process of creating their own computational narratives, and becoming self-directed in their future applications of computing

\section{The four learning modules completed up to this point}
As of this writing, four learning modules are complete, and shared on GitHub under CC-BY and BSD-3 licenses. 
The entry point repository is at \url{https://github.com/engineersCode/EngComp}. 
We have a fifth module under development, on the topic of Fourier analysis. 
Additional learning modules are planned, for supplementing core engineering courses. 
The first three modules have been tested in the classroom three times. 
Also, the first module was adopted by a team of data services librarians at the GW Library to conduct short, intensive Python workshops over three or four days. 
The fourth module (on linear algebra) was used once for leading a live tutorial at the SciPy Conference in July 2019 (the video of this tutorial is available at \url{https://youtu.be/4-P0gptDT40}), and once more in a classroom setting during Fall 2019 (with second-year engineering students).

\subsection*{Module 1: Get data off the ground with Python}
The first module assumes no prior coding experience, so the first three lessons are focused on creating a foundation with Python programming constructs, with minimal and basic mathematical content (to spare cognitive load). 
The fourth lesson introduces the basic data structure in scientific computing: arrays; the mathematics content is kept at high-school level.
The final lesson is a worked example of linear regression with real data, and contains some calculus-based content.

\begin{description}[style=unboxed]

\item[Lesson 1: Interacting with Python]---
Background: What is Python? Idea of interpreted vs. compiled language. Why use Python? It is a general-purpose and high-productivity language. Getting started: interactive Python (IPython). Using Python as a calculator. New concepts: function, string, variables, assignment, type, special variables (\texttt{True}, \texttt{False}, \texttt{None}). Supported operations, logical operations. Reading error messages.

\item[Lesson 2: Play with data in Jupyter]---
What is Jupyter? Working with Jupyter. Playing with Python strings: assignment, indexing, slicing. String methods: \texttt{count}, \texttt{find}, \texttt{index}, \texttt{strip}, \texttt{startswith}, \texttt{split}. Play with Python lists: assignment, nested lists, indexing, slicing. String methods: \texttt{append}, \texttt{index}. List membership. Iteration with \texttt{for}-statements. Conditionals.

\item[Lesson 3: Strings and lists in action]---
A full example using what you learned in lessons 1 and 2: playing with a text file containing the MAE Bulletin (list of courses with their numbers, description, pre-requisites). Reading a data from a file. Cleaning and organizing text data.

\item[Lesson 4: Play with NumPy arrays]---
Two of the most important libraries for scientific computing with Python: NumPy and Matplotlib. Importing libraries. NumPy functions to create arrays: \texttt{linspace}, \texttt{ones}, \texttt{zeros}, \texttt{empty}, \texttt{copy}. Array operations. Multidimensional arrays. Performance advantage of arrays over lists. Drawing 2D line plots of array data.

\item[Lesson 5: Linear regression with real data]---
A full worked example using real data of earth temperature over time. Step 1: reading data from a file. Step 2: plotting the data; making beautiful plots. Step 3: least-squares linear regression. Step 4: applying linear regression using NumPy. Split regression.

\end{description}

\subsection*{Module 2: Take off with Stats in Python}
This learning module builds from a foundation in Python programming to develop data practices and computational problem-solving. 
Students learn to handle data programmatically, reading data from files, cleaning and organizing data, and performing exploratory data analysis. 
They use real data, learn to make pretty data visualizations, and gain insight from data.

\begin{description}[style=unboxed]

\item[Lesson 1: Cheers! Stats with beers]---
Exploratory analysis using a data set of canned craft beers in the US. Introduces the \texttt{pandas} library and its data types: Data Frames and Series. Use \texttt{pandas} to read a data file, extract selected columns, and remove null values. Descriptive statistics: measures of central tendency and variability. Distribution plots: histograms with Matplotlib. Comparing with a normal distribution.

\item[Lesson 2: Seeing stats in a new light]---
Continuing with the data set of canned craft beers, this lesson focuses on visualizing statistics. For quantitative data: histograms and box plots; for categorical data: bar plots. Visualizing multiple data with scatter plots and bubble charts.

\item[Lesson 3: Lead in lipstick]---
A full worked example using what you learned in lessons 1 and 2: using data from studies by the US Food and Drug Administration on the lead content in lipstick, we fact-check alarming news headlines. Based on Prof. Kristin Sainani's lecture, ``Exploring real data: lead in lipstick,'' of her Stanford Online course ``Statistics in Medicine.''

\item[Lesson 4: Life expectancy and wealth]---
A deeper dive into \texttt{pandas} for data analysis, using data of life expectancy and gross-domestic product (income) per capita over time, for various countries across the world. Grouping data for analysis and dataframe manipulation.

\end{description}

\subsection*{Module 3: Tour the Dynamics of Change and Motion with Python}
This module builds from a foundation in Python programming to develop modeling and simulation practices, and computational problem-solving. 
Students learn to capture motion from images and videos, to compute velocity and acceleration from position data, to obtain velocity and position from accelerometer data, and to study differential models of mechanical vibrations.

\begin{description}[style=unboxed]

\item[Lesson 1: Catch things in motion]---
Working with images and videos in Python using \texttt{imageio}. Interactive Matplotlib figures in the notebook, and capturing mouse clicks on images for digitizing an object's position. Computing velocity and acceleration from position captures: a falling ball, and projectile motion. Computing numerical derivatives using differences. Free-fall acceleration from real data.

\item[Lesson 2: Step to the future]---
Computing velocity and position from accelerometer data: a roller-coaster ride. Using the \texttt{subplot()} function to draw more than one plot in the same figure. Euler's method for initial-value problems, and Taylor expansion showing first-order accuracy. The second-order differential model for an object in free fall written as two first-order differential equations, leading to a vector form. General design of a code to solve ordinary differential equations (ODEs). Application to free fall of a tennis ball and comparison with experimental data. Improved model accounting for air resistance.

\item[Lesson 3: Get with the oscillations]---
Differential model of a spring-mass system without friction: state vector and system in vector form. Amplitude growth with Euler's method on oscillatory systems, and the fix: Euler-Cromer method (semi-implicit Euler). Numerically observed order of accuracy using a convergence plot: numerical error with different time increments. Modified Euler's method, and observed order of accuracy.

\item[Lesson 4: Bird's-eye view of mechanical vibrations]---
General spring-mass systems with damping and a driving force, revealing a variety of behaviors. Presents a powerful new method to study dynamical systems based on visualizing direction fields and trajectories in the phase plane.

\end{description}

\subsection*{Module 4: Land on Vector Spaces with Python}
This module applies Python and core numerical libraries (NumPy, SymPy, Matplotlib) to explore the foundations of linear algebra, with a geometrical and practical approach. 
Students learn to view matrices as linear transformations of vectors, and develop intuition about their role in linear systems of equations. 
Playing with transformations, students understand eigenvalues and eigenvectors, and discover matrix decomposition. 
We use Python to compute all the eigenthings and apply them to population models in ecology, Markov Chains, and the Google Page Rank algorithm. 
Students learn about singular-value decomposition and its application to image compression, least squares problems, and linear regression.

\begin{description}[style=unboxed]
\item[Lesson 1: Transform all the vectors]---
What is a vector? The physicist's view versus the computer scientist's view. Fundamental vector operations: visualizing vector addition and multiplication by a scalar. Intuitive presentation of basis vectors, linear combination and span. What is a matrix? A matrix as a linear transformation mapping a vector in one space, to another space. Visualizing linear transformations. Matrix-vector multiplication: a linear combination of the matrix columns. Some special transformations: rotation, shear, scaling. Matrix-matrix multiplication: a composition of two linear transformations. Idea of inverse of a matrix as a transformation that takes vectors back to where they came from.

\item[Lesson 2: The matrix is everywhere]---
A matrix is a linear transformation: visualize it. Norm of a vector. A matrix maps a circle to an ellipse: visualize it. A vector that doesn't change direction after a linear transformation is an eigenvector of the matrix. A matrix is a system of equations: visualize it (row perspective). Inconsistent and underdetermined systems. A matrix is a change of basis: visualize it. An inverse of that matrix will change the vector's coordinates back to the original basis. Matrices in three-dimensional space: linear transformations in 3D; 3D systems of linear equations; dimension and rank. Visualize the transformations of rank-deficient matrices.

\item[Lesson 3: Eigenvectors for the win]---
Geometry of eigendecomposition. Eigenvectors revisited: a matrix transforms a circle to an ellipse, whose semimajor and semiminor axes align with the eigenvectors. Composition of scaling transformation and a rotation transformation: not enough! Complete the composition. Symmetric matrices, orthogonal eigenvectors. Eigendecomposition in general. Diagonalizable matrices. Similar matrices. Eigendecomposition is similarity via a change of basis. Compute eigenthings in Python, using NumPy or SymPy. Eigenvalues in ecology: matrix population models. Markov chains. PageRank algorithm.

\item[Lesson 4: Stick to the essentials: SVD]---
Geometrical interpretation of singular value decomposition (SVD). While eigendecomposition is a combination of change of basis and stretching, SVD is a combination of rotation and stretching, which can be treated as a generalization of eigendecomposition. Example: SVD in image compression. A 2D image can be represented as an array where each pixel is an element of the array. By applying SVD and dropping smaller singular values, we can reconstruct the original image at a lower computational and memory cost. Non-square matrices: SVD in general; pseudo-inverse. Application to linear least squares; linear regression with SVD.

\end{description}

\section{Hybrid to online with Jupyter-first course design}

``Jupyter first'' alludes to the idea of developing a course first as a set of Jupyter notebooks, then building both an online course and an on-campus learning experience based on those notebooks. 
For the \emph{Engineers Code} series, I am creating an online mini-course for each module,  in the format of a ``massive open online course'' (MOOC), using the Open edX course platform. 
The site is found at \url{http://openedx.seas.gwu.edu}, and anyone can register and enroll in the online courses, at no cost. 

Open edX is a full-featured open-source platform for online courses, used by millions of learners via the edX consortium, large MOOC platforms abroad (France, China, Spain, and others), and institutional deployments. 
It allows for third-party extensions with its XBlock specification. 
With outside technical partners, we developed ways to integrate both content and assessments based on Jupyter into an online course in Open edX. 
We have contributed two XBlocks to build courses based on Jupyter: the Viewer, and a Jupyter Grader for auto-graded student assignments (both released as open source). 
With the Jupyter Viewer Xblock, a course designer can build learning sequences with content pulled dynamically from a public Jupyter notebook (e.g., on a GitHub repository). 
Jupyter-first courses can be written using an open development model (like any open-source software project), collaboratively and under version control. 
Once the material is ready, the course builder can create a MOOC-style course in Open edX, pulling the content from the notebooks without duplication in the course platform. 
(Note that Open edX has no concept of version control.) 
One can interleave short videos and graded sub-sections using the built-in problem types, or using the Graded Jupyter XBlock. 
Our course development workflow is the product of several years of refinement, and applies evidence-based instructional design. 
Combined with modern pedagogies used in the classroom, like active learning via live coding, an instructor can create learning experiences that are effective on campus and online. 

\bigskip

\begin{figure*}
\begin{center}
\includegraphics[width=0.8\textwidth]{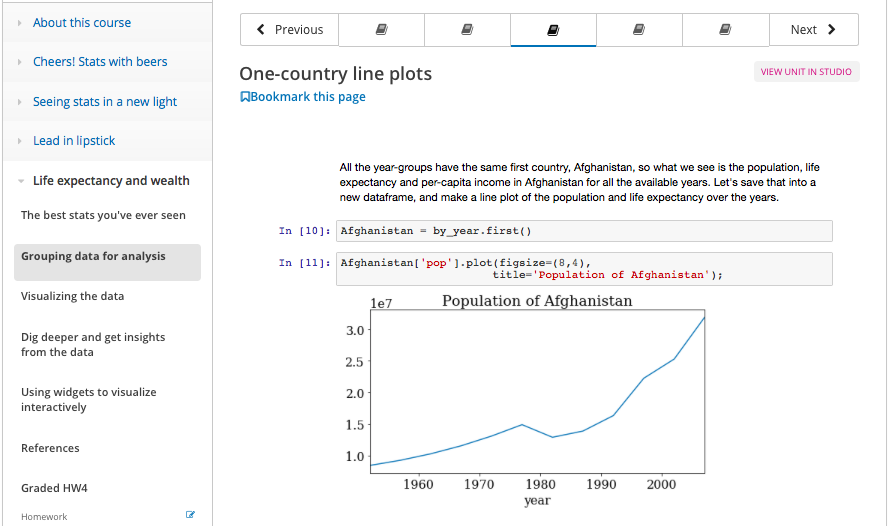}
\caption{Screenshot of Jupyter-based content (from Module 2) displayed in an Open edX learning sequence, showing rich display of code and output.\label{fig:openex}}
\end{center}
\end{figure*}

\noindent \textbf{Summary of the Jupyter XBlocks:}
Following is a short description of an instructor's experience building a Jupyter-based course in Open edX using our XBlocks:
\emph{Jupyter Notebook Viewer XBlock:} From any public Jupyter notebook (e.g., in a public repository on GitHub), pull content into a course learning sequence using the notebook URL (dynamic content). 
Use optional start and end marks (any string from the first cell to include, and the first cell to exclude) to break a long notebook into unit-sized parts. 
This allows course authors to develop their course content as Jupyter notebooks, and to build learning sequences reusing that content, without duplication. 
It also has the added benefit that the development of the material can be hosted on a version-controlled repository. (Open edX, itself, doesn't provide version control of course content.)
Moreover, the full richness of presentation in a Jupyter notebook's rendering is available for display inside the course: formatted equations, syntax-highlighted code, output from computations, data visualizations, etc. (See Figure \ref{fig:openex}.)
The code repository for the XBlock is at \url{https://github.com/ibleducation/jupyter-viewer-xblock}, and is open source under a BSD-3 license.

\emph{Graded Jupyter Notebook XBlock:} An instructor creates an assignment using the \texttt{nbgrader} Jupyter extension \cite{hamrick2016,blank2019}, then can insert a graded sub-section in Open edX that will deliver this assignment (as a download), auto-grade the student's uploaded solution, and record the student's score in the gradebook. 
The XBlock instantiates a Docker container with all the required dependencies, runs \texttt{nbgrader} on the student-uploaded notebook, and displays immediate feedback to the student in the form of a score table. 
The code repository for the XBlock is at \url{https://github.com/ibleducation/jupyter-edx-grader-xblock}, and is open source under a BSD-3 license.

\bigskip 

Note that the course platform displays learning sequences with Jupyter-based content, and can auto-grade assignments made in Jupyter, but it does not provide interactive computing in cloud resources. 
Our Open edX platform is open to anyone for registration, and it is not economically feasible for us to offer cloud computing to the public. 
Students registered in our regular on-campus courses, or attending the Python camps in the Library, have access to a Jupyter Hub instance, for all their interactive computing needs. 
Learners following our courses from around the world can install Jupyter in their personal computers, or use one of the public cloud Jupyter services, like Binder (free, by Project Jupyter), CoCalc (commercial, with a free tier), or others.

\section{Evaluation}

We have some limited evaluation to report, based on student surveys conducted independently of the instructor, that targeted not the effectiveness of the learning modules but rather the impact they may have in student attitudes to open culture. 
(Our application to the Institutional Review Board, IRB, in the Fall 2017 received an exempt determination.)
During the first semester-long course using the (first three) learning modules, we carried out a qualitative study on how a new OER medium, Jupyter Notebooks, may (or may not) impact attitudes of undergraduate engineering students toward sharing and openness. 
Various research efforts in the last decade have focused on the impact on student outcomes from using OER. 
We were interested in whether our use of Jupyter, a new genre for OER, may also influence students in their attitudes and capacities for collaboration, community involvement, and open practices. 
During classroom discussions, we often emphasized the open-source nature of the tools we used, and promoted the idea that students could adopt these tools for any initiatives, during and after their studies. 
Out of a cohort of 52 students that semester (Fall 2017), 16 voluntarily answered a survey, and six agreed to be interviewed by a member of staff in the GW eDesign unit. 
The survey asked students their opinion about the online nature of the course materials, compared with traditional textbooks: 9 out of 16 students said they preferred the online presentation, 3 said they liked it less (3 had no preference, one did not answer). 
The pre-survey also aimed to sample a baseline of attitudes toward sharing and openness, via an open-ended question. 
Students' answers reflected equally pro-sharing and anti-sharing attitudes, the latter mostly referring to others taking ``credit'' for their work. 
Certain cultural barriers persist that inhibit openness, it seems: some students' expectation of having a textbook, and a culture of competition and fixation on grades. 
We concluded that a single semester-long course is not enough time to make a change in culture. 
Open educational practices should be threaded through several courses, and start early. 
Collaboration with other instructors is needed to reinforce ideas, skills, attitudes across more than one course.

In the Spring 2019 semester, the introductory modules were taught as a new first-year course (as part of a two-course series).
Quantitative evaluation with this cohort targeted students' perceived gains in computing skills, and changes in attitudes towards coding. 
Out of 48 students taking the course, 23 responded to a survey (69\% male and 31\% female). 
At the end of the course, the responses to ``How prepared are you for learning to code? '' increased from an average of 5.4 to 7.4 (on a 0--10 scale). 
The survey also asked several questions on perceptions about using coding skills after graduation, most of which displayed a modest increase. 
On the prompt ``I plan to use coding in my career after graduation,'' the average of responses increased from 6.2 to 7.3. 

We also carried out a short survey on perceived usefulness, targeting the first cohort of students that used the learning modules in Fall 2017, when they were in their second year. 
They are Seniors now, and have the benefit of time in assessing the usefulness of the materials.  
The survey just had three questions on a scale from 1 to 5 (from ``not at all'' to ``a lot''), as follows:
\begin{enumerate}
\item Do you find yourself using these computational tools in other courses later? 
\item Have you learned more Python techniques over time?
\item Do you think you will use Python/Jupyter/computing-in-general in your future career?
\end{enumerate}
\noindent Thirty-two out of 50 possible students responded (two students had left the university) this survey. 
To the first question (used the computational tools in other courses later), 78 percent chose 5 (``a lot'') and an additional 19 percent chose 4.
Only one student chose 1 (``not at all''). 
To the second question (learned more over time), 37\% chose 5, and 41\% chose 4.

The evaluation is modest, but our focus has not been to conduct education research, but rather to create the learning materials according to our design specifications, and to develop complementary technology (Jupyter viewer and auto-grader XBlocks) to provide a blended learning experience.

\section{Conclusion}

This article for the \emph{CiSE} Special Issue on Computational Science and Engineering Education presents our collection of learning modules for undergraduate engineering students, aimed at developing foundational computing and data skills. 
These modules are developed in the open-source model (on GitHub), designed to be reused, and shared under standard public licenses allowing redistribution and revision by users to fit their needs. 
Our long-term vision is to collaboratively develop several more modules, aiming to populate the engineering curriculum with computational content. 
Ultimately, the education of STEM graduates can be indelibly transformed by computing and data skills becoming \emph{infrastructural}, when all learners are led and supported to become proficient in them, and to apply them in authentic technical contexts. 
This is the defining feature of a \emph{literacy}: a socially widespread deployment of skills and capabilities that become material support to achieve valued intellectual ends \cite{disessa2001}.
Knowing how to read and write, that is, the conventional meaning of literacy, is not only highly valued in society, but the bedrock of all education. 
Similarly, computing and data skills are today indispensable to almost all STEM fields, and form the basis of a new literacy.

The \emph{Engineers Code} project implements key concepts and design principles, distilled from several years developing instructional materials using Python and Jupyter, and informed by the education literature.  
It is not education research, but rather an implementation project. 
The education-research literature supports our design principles, however.
Active learning is superior to lectures in both comprehension and recollection \cite{freeman2014active} and we apply it using Jupyter notebooks combined with live coding and other tactics like pair programming. 
The worked-example effect \cite{chen2015worked}, the most-studied cognitive-load effect, explains why our design helps students manage the complexity of learning applied computing.
Modularization, chunking, interleaving: these are known effective techniques in learning, which we have put to work in this project. 
Ongoing evaluation is helping us confirm the positive effect on students, and build confidence to share with others interested in adopting our approach. 

The project is also producing online `mini-courses' using a MOOC platform (Open edX).
In the process, I have refined an approach that I call \emph{Jupyter-first} course development, where a course is first written as a set of Jupyter notebooks, and the learning sequence in the online course platform pulls from this content dynamically. 
To achieve it, we created with technical partners the third-party extension (XBlock) to embed content from a public Jupyter notebook in a course.
We also created an extension to allow auto-grading of student assignments written as Jupyter notebooks. 
All this technology has been released to the public and is open source. 
The online modules are also open for anyone to enroll and follow at their own pace, and could even be assigned by other instructors to complement their courses. 
We know of one instructor at another university who did so, and also have collaborated with the GW Libraries to help them offer Python camps using a combination of our first module and parts of the second. 
They condensed the face-to-face learning to 3 or 4 days, and learners complete the auto-graded assignment in the online platform (Open edX) after the camp to receive a certificate of completion. 
These Python camps are in high demand, and the GW Libraries reported that registrations fill all available seats in under 8 hours from announcing the event! 
Three camps were held in 2019, and the Library plans to continue offering them in 2020. 
We have also started to support these informal learners, and the learners in the official GW courses, via Study Hall sessions in the Library, staffed by undergraduate student tutors. 
Our vision is to help coalesce an active learning community in the university for learning foundations of programming in applied contexts. 

\section*{Acknowledgements}

This work received support from grant \#1730170  of the Office of Advanced Cyberinfrastructure (OAC), National Science Foundation. 
The student surveys and interviews during Fall 2017 were conducted by Tara Lifland, then an instructional designer at the  George Washington University eDesign Shop and \href{http://openedgroup.org/fellowship}{OER Research Fellow} of the Open Education Group. 
The surveys during the Spring 2019 were conducted by Prof. Ryan Watkins, from the GW School of Graduate Education and Human Development. 
The GW Libraries Python Camps are led by Megan Potterbusch and Laura Wrubel.
Several graduate students in Barba's group have contributed to her educational initiatives, and the development of learning modules (for this project and its predecessors): Natalia C. Clementi, Pi-Yueh Chuang, Gilbert Forsyth, Olivier Mesnard, and Tingyu Wang.
With special and heartfelt gratitude towards all the contributors to Project Jupyter and its ecosystem of tools for education. 
Project Jupyter is a fiscally sponsored project of NumFOCUS (\href{https://numfocus.org}{https://numfocus.org}), a 501(c)(3) US non-profit with a mission to promote open practices in research, data, and scientific computing.

\bigskip

\begin{wrapfigure}{r}{0.33\linewidth}
\includegraphics[width=3 cm]{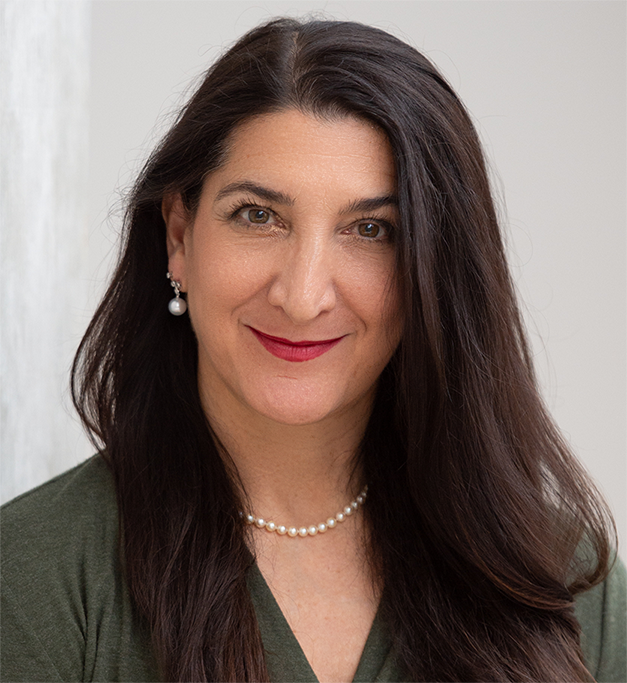}
\end{wrapfigure}
\noindent \textbf{Lorena A. Barba} is a professor of mechanical and aerospace engineering at the George Washington University. Her research interests include computational fluid dynamics, biophysics, and high-performance computing. She is co-Editor of the CiSE Reproducible Research Track, Associate Editor for The ReScience Journal, Associate Editor-in-Chief of the Journal of Open Source Software, and Editor-in-Chief of the Journal of Open Source Education. Barba received a PhD in aeronautics from the California Institute of Technology. Contact her at labarba@gwu.edu and find her website at \href{http://lorenabarba.com}{http://lorenabarba.com}.

\bibliographystyle{IEEEtran}
\bibliography{comp_edu}
%

\end{document}